# Comparative Study of Lateral and Vertical β-$Ga_2O_3$ Photoconductive Switches via Intrinsic and Extrinsic Optical Triggering

Vikash K. Jangir, Graduate *Student Member, IEEE,* and Sudip K. Mazumder, *Fellow, IEEE*

***Abstract***—**Gallium oxide ($Ga_2O_3$), with its ultra-wide bandgap (~4.8 eV) and high breakdown field (~8 MV/cm), is a leading candidate for photoconductive semiconductor switch (PCSS) in high-power and high-speed pulsed applications. This work, for the first time, presents a systematic experimental comparison of lateral and vertical β-$Ga_2O_3$ PCSS under both intrinsic (245 nm) and extrinsic (280 nm, 300 nm, and 445 nm) optical excitation. Under intrinsic excitation, where carrier generation is confined near the surface due to the shallow absorption depth (~0.1–1 µm), the lateral PCSS demonstrated higher photocurrent performance compared to the vertical structure. In contrast, under extrinsic excitation, which enables deeper penetration into the bulk, the vertical PCSS exhibited enhanced switching performance due to a more uniform electric-field distribution across the device volume. These results highlight the critical role of device geometry and carrier generation mechanism in optimizing $Ga_2O_3$ PCSS performance and provide valuable guidance for developing efficient and cost-effective high-voltage PCSS devices.**

*Index Terms*- **Intrinsic, electric field, extrinsic, Fe-doped $Ga_2O_3$, laser, lateral, optical triggering, photoconductive semiconductor switch (PCSS), photo response, vertical**

## I. INTRODUCTION

GALLIUM oxide ($Ga_2O_3$) is an emerging ultra-wide bandgap semiconductor (~4.8 eV) with a high theoretical breakdown field (~8 MV/cm) [1][2], making it highly attractive for high-power electronic and optoelectronic applications. These intrinsic properties position $Ga_2O_3$ as a promising material for photoconductive semiconductor switches (PCSS), which demand high voltage hold-off and fast, optically triggered switching capabilities [3].

Device geometry and optical-excitation methods play critical roles in determining PCSS performance. Lateral geometry, where the anode and cathode are on the same surface, yield simpler devices to fabricate; however, these structures are often limited by surface flashover, where electric field crowding at the electrode edges can lead to premature breakdown. To mitigate this, vertical PCSS designs—where electrodes are placed on opposite sides of the substrate—have been explored. Vertical configurations offer advantages such as improved field uniformity, reduced edge field enhancement, and enhanced current spreading area, resulting in superior breakdown performance and operational reliability[4][5][6].

The excitation mechanism—intrinsic or extrinsic—also strongly affects switching behavior. For $Ga_2O_3$, intrinsic excitation, based on band-to-band transitions, requires deep-ultraviolet (DUV) light sources such as 245-nm lasers. Such DUV lasers are typically bulky, inefficient, and prohibitively expensive, posing a significant barrier to the widespread, practical deployment of $Ga_2O_3$ PCSS devices. In contrast, extrinsic excitation of PCSS leverages sub-bandgap light absorption via trap states, enabling the use of compact and cost-effective light sources such as 445-nm diode lasers. This makes extrinsic triggering appealing for scalable PCSS applications.

Defect states such as intentionally doped iron (Fe) and unintentionally introduced iridium (Ir) during growth, facilitate sub-bandgap absorption in $Ga_2O_3$ [7][8]. Fe forms deep acceptor levels ~0.7–0.8 eV below the conduction band and can emit electrons under sub-bandgap illumination. Ir-related mid-gap states (~2.2–2.3 eV) can also contribute to carrier generation under visible-light excitation.

This work, for the first time, presents a systematic experimental comparison between lateral and vertical β-$Ga_2O_3$ PCSS devices, directly evaluating their performance under both intrinsic and a wide range of extrinsic optical-triggering conditions. By correlating the experimental switching results with detailed simulations of the internal electric field and wavelength-dependent photogeneration profiles, this study provides critical insights into how the interplay between device geometry and excitation mechanism dictates device performance.

## II. DEVICE STRUCTURE AND EXPERIMENTAL SETUP

A Fe-doped semi-insulating β-$Ga_2O_3$ substrate (Fe ~ $1 \times 10^{17}$ $cm^{-3}$) obtained from Kyma Technologies was used to fabricate both lateral and vertical PCSS structures. The schematic cross-section and top-view micrographs of the fabricated lateral and vertical devices are shown in Fig. 1 and Fig. 2, respectively. For the lateral PCSS, Rogowski-profile contacts [9] were employed to enhance electric-field uniformity across the gap between the contact electrodes. The electrodes consist of a Ti (50 nm)/Au (100 nm) metal stack deposited via e-beam evaporation, followed by rapid thermal annealing at 480 °C for 1 minute in a nitrogen ambient to establish ohmic contact [10]. A standard lift-off process using AZ 1518 positive photoresist was used for electrode patterning. The device features a 450 µm electrode gap and a 1950 µm contact length, as illustrated in Fig. 1(b).

The information, data, or work presented herein was funded in part by the Advanced Research Project Agency-Energy (ARPA-E), U.S. Department of Energy, under Award Number DE-AR0001879. The views and opinions of authors expressed herein do not necessarily state or reflect those of the United States Government or any agency thereof.

The authors are with the Department of Electrical and Computer Engineering, University of Illinois Chicago, Chicago, IL 60607 USA (e-mail: vjangi3@uic.edu; mazumder@uic.edu).

For the vertical PCSS, a uniform Ti (50 nm)/Au (100 nm) metal contact was deposited on the bottom surface of the substrate. The top contact was patterned with alternating metal and transparent regions to enable light injection and achieve a more uniform vertical electric field, as shown in Fig. 2(b). The deposition and patterning steps were identical to those used for the lateral structure. The effective electrode area of the vertical switch is 0.01 cm², while that of the lateral switch is 0.0088 cm². After annealing, the ohmic contacts for both designs were verified using I-V curves as shown in Fig. 3. Both the lateral PCSS and the vertical PCSS were packaged in a TO-257 metal can package for electrical pulse characterization.

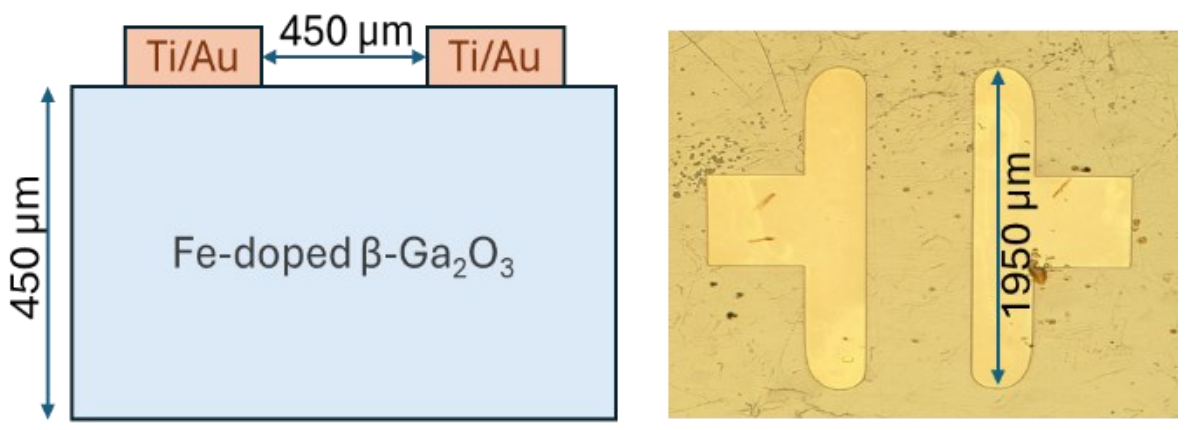

(a) (b)

Fig. 1. (a) Schematic cross-section of lateral $\beta$-$Ga_2O_3$ PCSS. (b) Top view of the fabricated device.

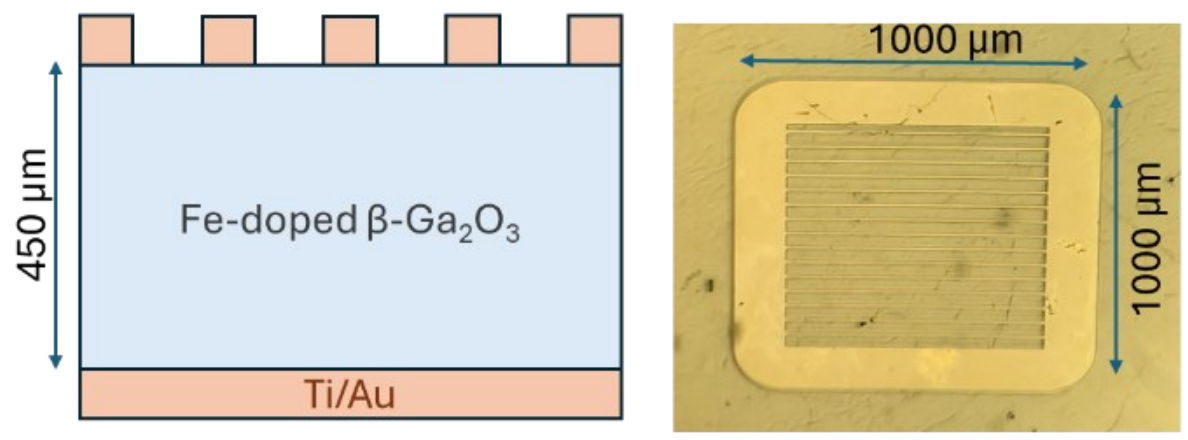

(a) (b)

Fig. 2. (a) Schematic cross-section of vertical $\beta$-$Ga_2O_3$ PCSS. (b) Top view of the fabricated device.

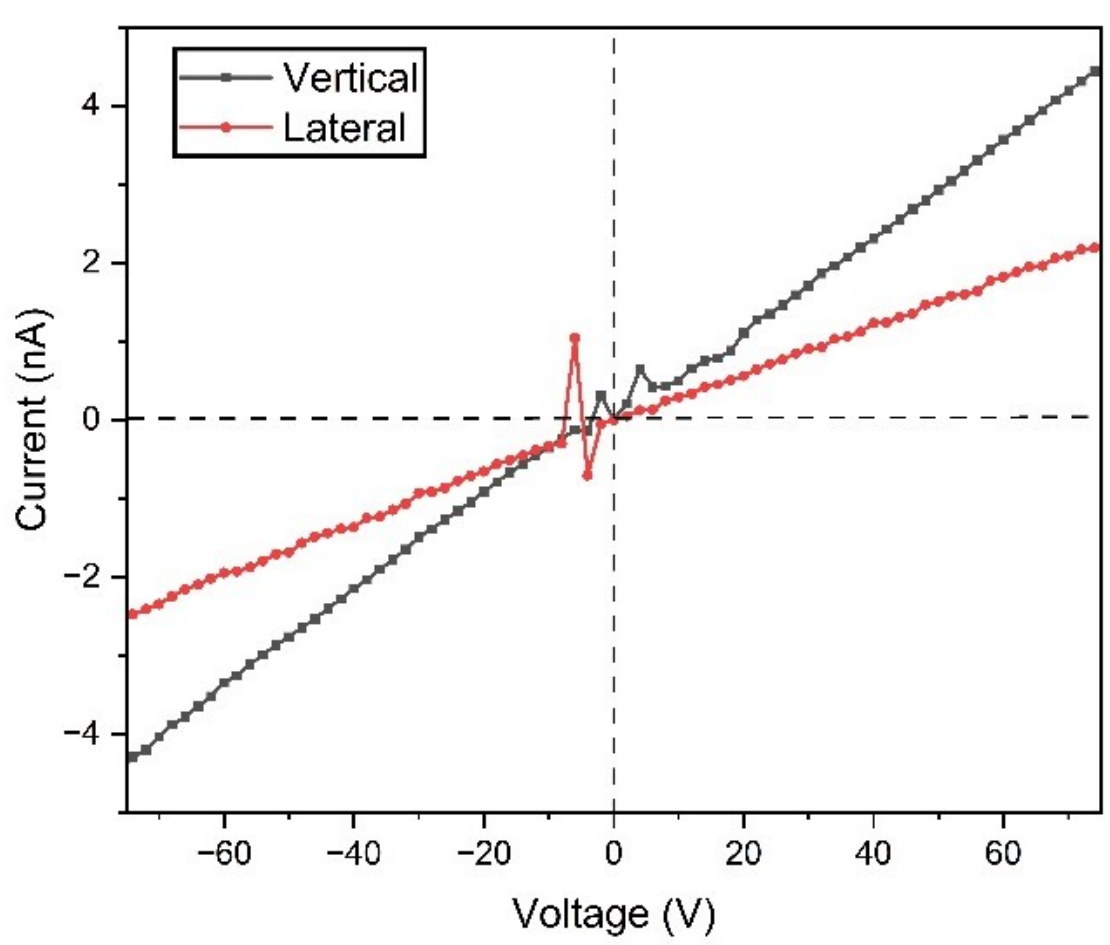

Fig. 3. I-V characteristic for ohmic contact

Fig. 4 shows the experimental setup used for optical triggering of the two switches. A 100 nF capacitor is charged to the desired DC bias through a 1 kΩ current-limiting resistor. Both the lateral and vertical PCSS devices are triggered using an OPOLETTE UX06230U tunable laser with a beam diameter of 3 mm and a full width at half maximum (FWHM) of 7 ns. Owing to the large beam diameter relative to the active area of the PCSS, only about 15% of the incident optical power was effectively coupled into the device. The resulting electrical pulse is measured across a 50 Ω load resistor using a Tektronix TPP1000 voltage probe (1 GHz bandwidth) and a Tektronix MSO46 oscilloscope (1 GHz bandwidth). The laser pulse profile is simultaneously monitored using a high-speed silicon photodetector (DET10A2, Thorlabs).

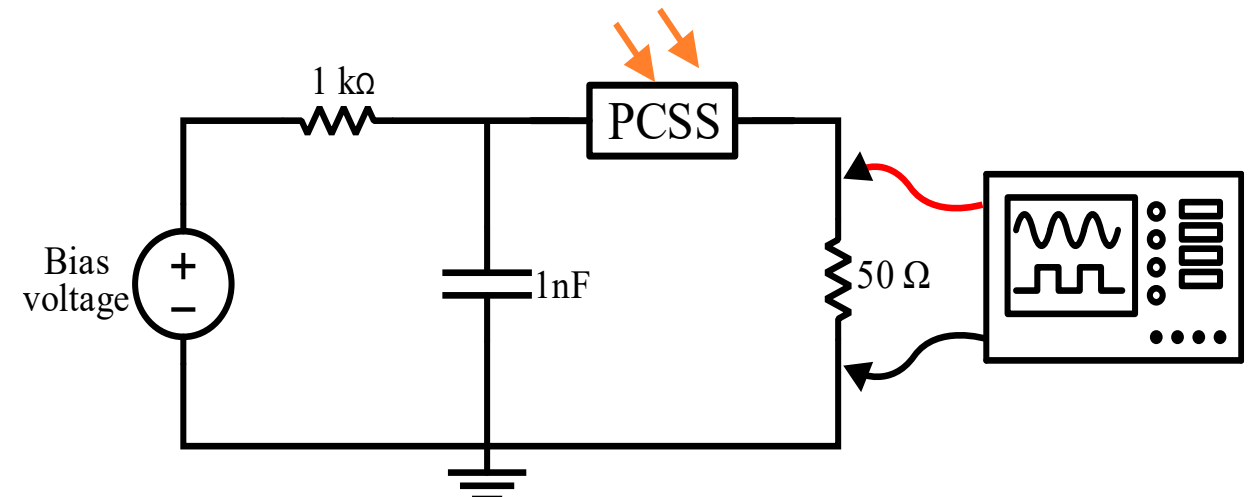

Fig. 4. Schematic of the experimental setup for testing the $Ga_2O_3$-based lateral and vertical devices.

## III. RESULTS AND DISCUSSION

To experimentally compare the switching characteristics of the lateral and vertical $Ga_2O_3$ switches, they were evaluated under both intrinsic (245 nm) and extrinsic (280 nm, 300 nm, and 445 nm) optical excitations. The temporal waveforms of the optical trigger pulses and the corresponding device currents, measured across a 50 Ω load resistor, are presented in Fig. 5.

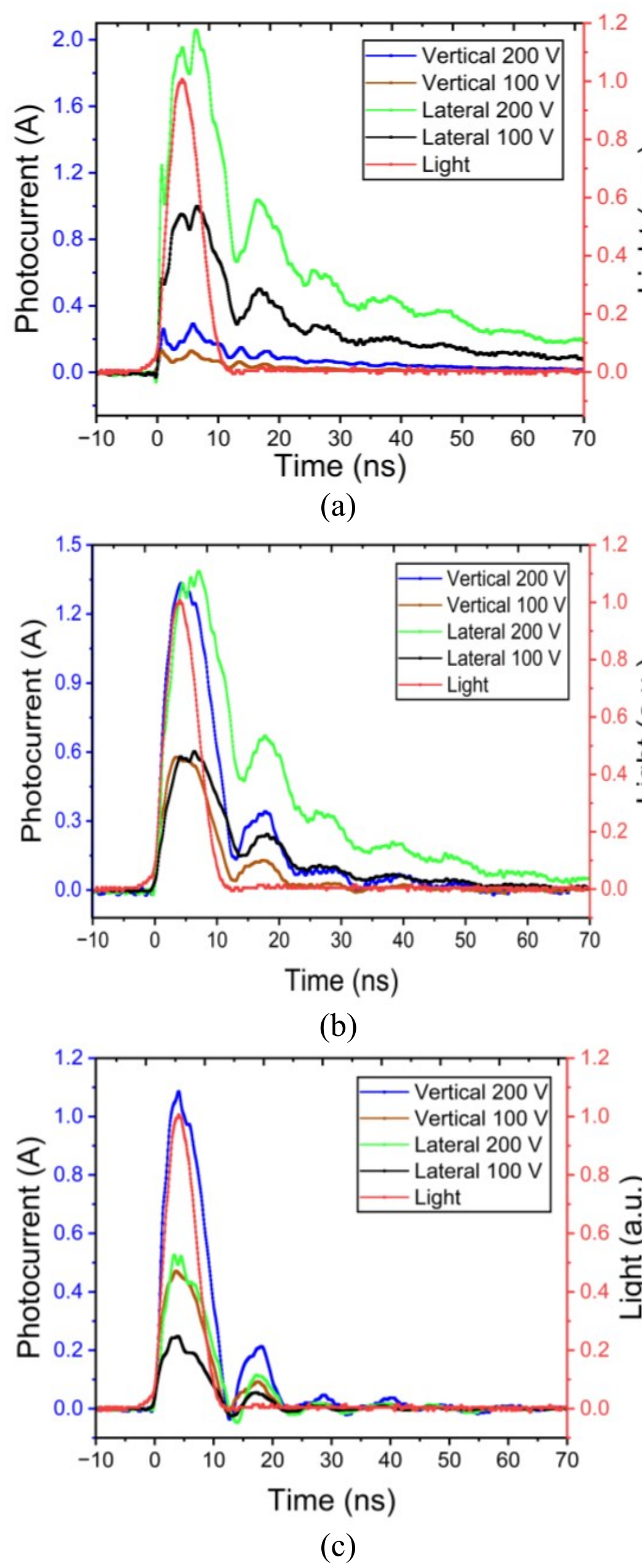

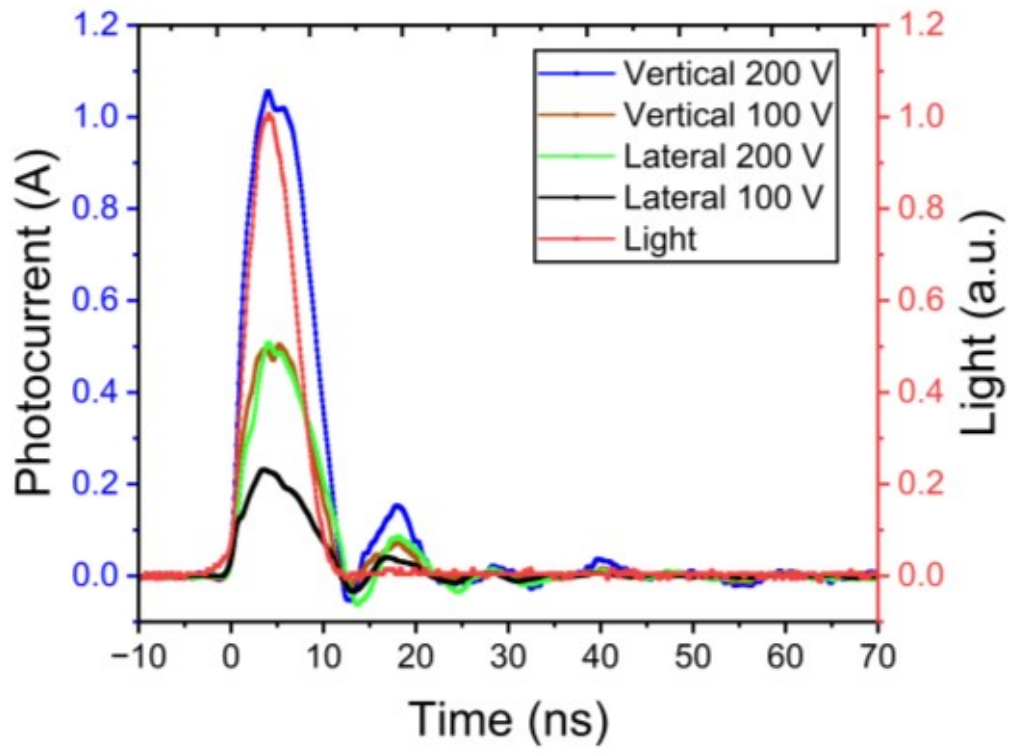

(d)

Fig. 5. Experimental optical trigger pulse waveform (red) and the corresponding output current waveforms of the lateral PCSS and the vertical PCSS obtained under different excitation conditions: (a) Intrinsic excitation using a 245-nm laser with 6.07 × $10^5$ W/cm$^2$ pulse power density; (b) extrinsic excitation using a 280-nm laser with 6.07 × $10^5$ W/cm$^2$ pulse power density; (c) extrinsic excitation using a 300-nm laser with 6.07 × $10^5$ W/cm$^2$ pulse power density; and (d) extrinsic excitation using a 445-nm laser with a higher pulse power density of 5.66 × $10^6$ W/cm$^2$

For intrinsic excitation, shown in Fig. 5(a), the lateral and the vertical devices were illuminated with a 6.07 × $10^5$ W/cm$^2$ laser pulse of wavelength 245 nm under bias voltages of 100 V and 200 V. For extrinsic excitation, 6.07 × $10^5$ W/cm$^2$ at wavelengths of 280 nm and 300 nm, and a 5.66 × $10^6$ W/cm$^2$ pulse at a wavelength of 445 nm, were used under the same bias conditions. The response to the 280 nm pulse is shown in Fig. 5(b), the 300 nm pulse in Fig. 5(c), and the 445 nm pulse in Fig. 5(d). A summary of the key performance parameters for the lateral PCSS and the vertical PCSS under all excitation conditions is provided in Table 1. The minimum on-state resistance ($R_{ON}$) of the PCSS is calculated as $R_{ON} = V_{DC} / I_{peak} - R_L$, where $V_{DC}$ is the DC bias voltage, $I_{peak}$ is the peak photocurrent measured across the load resistor $R_L$, and $R_L$ = 50 Ω.

The experimental results reveal a strong dependence on both the device architecture and the optical-excitation wavelength. During intrinsic excitation (245 nm), the lateral PCSS yields a better photoresponse than the vertical PCSS, as shown in Fig. 5(a). As detailed in Table 1, the lateral device achieved an on-resistance ($R_{on}$) of just 54.17 Ω, compared to 719.23 Ω for the vertical device, and its responsivity was over eight times higher (1.25 × $10^{-4}$ A/W vs. 1.41 × $10^{-5}$ A/W).

However, this performance gap diminishes under near-sub-bandgap illumination at 280 nm, where the responses of the two devices become comparable [Fig. 5(b)]. Similar photocurrent magnitudes arise from the interplay between photogeneration depth and electric-field distribution. At this wavelength, carriers are generated up to ~150 μm deep; in the lateral device, the field strength decays rapidly with depth, reducing extraction efficiency, whereas the vertical device sustains a uniform bulk field that enables effective carrier collection throughout the illuminated region. At deeper sub-bandgap excitations, a clear performance crossover is observed, with the vertical PCSS outperforming the lateral structure at both 300 nm and 445 nm [Fig. 5(c) and 5(d)]. As shown in Table I, at 445 nm the vertical device exhibits an $R_{on}$ of 140.47 Ω—less than half that of the lateral device (350 Ω)—while delivering nearly twice the responsivity.

It is particularly noteworthy that while the lateral device's responsivity peaks at 245 nm and continuously declines with increasing wavelength, the vertical device's responsivity is highest in the 280–300 nm range.

This demonstrates that the vertical geometry is not only superior for deep sub-bandgap excitation but is optimally configured for the entire range of extrinsic triggering investigated here. This contrast arises from differences in absorption depth, electric-field distribution, and carrier collection efficiency between the two geometries. These distinctions are corroborated by the optical transmittance measurements (Fig. 6), the electric field simulations (Fig. 7), and the photogeneration profiles (Fig. 8).

The simulated quiescent (static) electric field distributions for lateral and vertical $Ga_2O_3$ geometries under applied DC bias, in the absence of optical excitation, are presented in Fig. 7. The lateral structure exhibits a highly non-uniform field, which is concentrated at the surface near the electrode edges [Fig. 7(a)] and, as quantified by the vertical line-cut in Fig. 7(b), decays exponentially into the substrate. In stark contrast, the vertical geometry [Fig. 7(c)] establishes a uniform electric field throughout the bulk drift region, confining the peak fields to the immediate anode and cathode interfaces, as confirmed by its corresponding line-cut [Fig. 7(d)].

TABLE I.

Summary of the experimental switching performances of the lateral PCSS and the vertical PCSS obtained under intrinsic and extrinsic excitations and at a bias of 200 V

| **Structure** | **Trigger wavelength (nm)** | **$R_{OFF}$ (GΩ)** | **$R_{ON}$ (Ω)** | **$R_{OFF}$ /$R_{ON}$** | **Normalized Responsivity (A-cm/kV-W)** |
|---|---|---|---|---|---|
| Lateral | 245 | 21.8 | 54.17 | 4.02 × $10^8$ | 8.2 × $10^{-5}$ |
| | 280 | 21.8 | 100.37 | 2.17 × $10^8$ | 5.58 × $10^{-5}$ |
| | 300 | 21.8 | 334.61 | 6.51 × $10^7$ | 2.2 × $10^{-5}$ |
| | 445 | 21.8 | 350 | 6.23 × $10^7$ | 2.26 × $10^{-6}$ |
| Vertical | 245 | 12 | 719.23 | 1.67 × $10^7$ | 9.27 × $10^{-6}$ |
| | 280 | 12 | 94.92 | 1.26 × $10^8$ | 5.12 × $10^{-5}$ |
| | 300 | 12 | 135.18 | 8.88 × $10^7$ | 4.01 × $10^{-5}$ |
| | 445 | 12 | 140.47 | 8.54 × $10^7$ | 4.76 × $10^{-6}$ |

This difference in field distribution explains the geometry-dependent performance. At 245 nm, $Ga_2O_3$ exhibits strong absorption near the surface with an absorption coefficient of ~6.0 × $10^4$ cm$^{-1}$ [11][12], generating carriers in the high-field region of the lateral device.

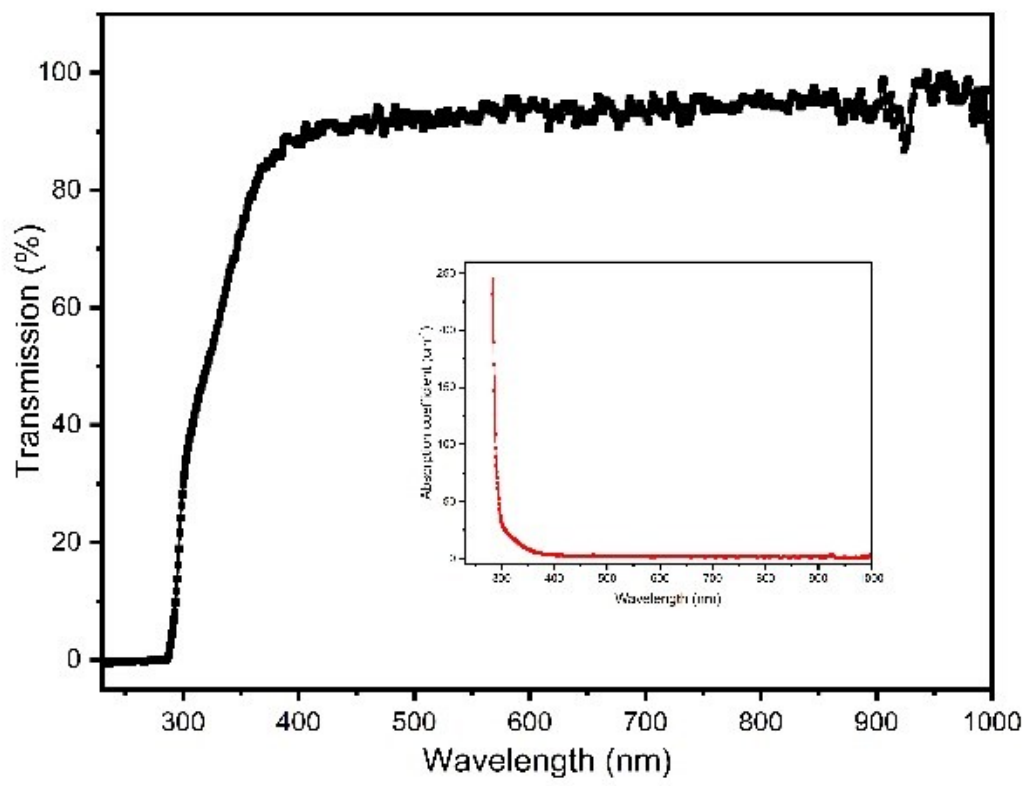

Fig. 6. Experimental optical transmission spectra of the Fe-doped β-$Ga_2O_3$

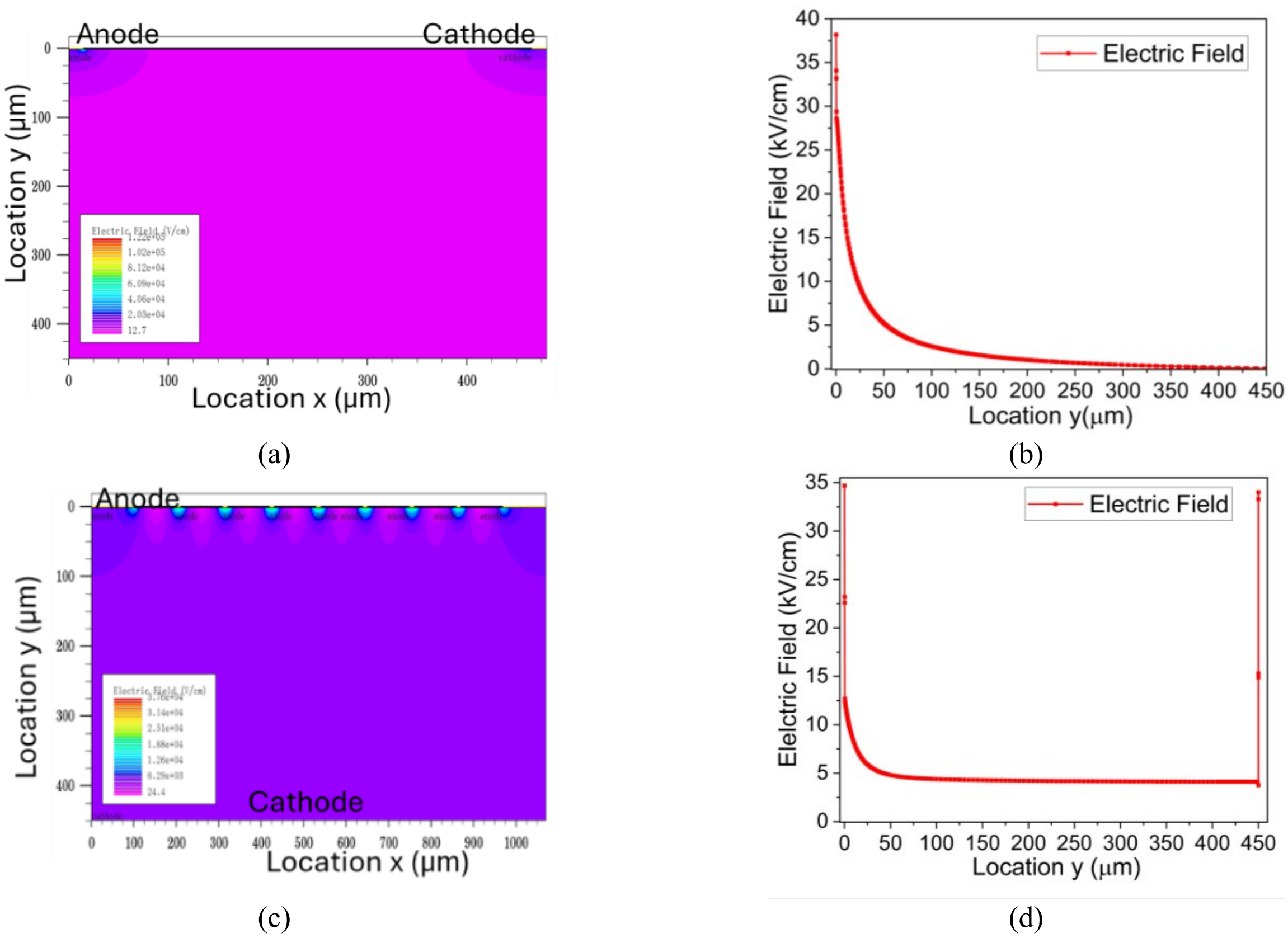

Fig. 7. TCAD Silvaco based simulation results: (a) Electric-field contour plot for the lateral device; (b) Vertical 1-D electric field profile for the lateral device; (c) 2D electric-field contour plot for the vertical device; and (d) Vertical 1-D electric field profile for the vertical device.

This excellent overlap between carrier generation [Fig. 8(a)] and the strong surface electric field [Fig. 7(a)] allows for efficient carrier collection and a large photocurrent. As the excitation wavelength increases, light penetrates deeper into the bulk. However, for longer wavelengths such as 300 nm and 445 nm, the photogeneration becomes more uniform and occurs deep within the substrate [Figs. 8(c) and 8(d)]. For the lateral geometry, these deep carriers are created in a region where the electric field has diminished to less than 5 kV/cm at depths beyond 50 µm. This weak field leads to poor carrier extraction and a significantly reduced photocurrent. Conversely, this deep and uniform photogeneration profile is an ideal match for the uniform bulk electric field of the vertical device [Fig. 7(c)]. The strong, constant field throughout the vertical device's drift region allows for efficient collection of these deep-generated carriers, explaining its superior performance at longer excitation wavelengths. Furthermore, both lateral and vertical $Ga_2O_3$ PCSS devices exhibit long trailing photocurrent under 245 nm and 280 nm excitation. Cassandra *et al.* reported sharp PL emission peaks at 690–709 nm, strong photoluminescence excitation (PLE) absorption bands at 430 nm and 600 nm, and a band edge onset near 280 nm [13]. Complementary photo-EPR studies have shown that optical excitation above ~2.9 eV promotes valence-band-to-gallium-vacancy ($V_{Ga}$) transitions that leave behind self-trapped holes (STHs) on oxygen and demonstrated that $V_{Ga}$ does not interact directly with Fe or Ir impurities [14]. Strong band-to-band absorption at 245 nm and near-edge absorption at 280 nm therefore generate conduction-band electrons, while the corresponding holes rapidly localize as STHs at ~4.4 eV below the conduction band [15–18]. These immobile STHs act as deep localized centers that suppress electron–hole recombination, enabling long-lived band-to-band conduction electrons that give rise to the trailing current, or persistent photoconductivity (PPC), observed under 245 nm and 280 nm excitation [Figs. 5(a) and 5(b)]. In contrast, 300 nm and 445 nm excitation lie below the absorption edge and primarily involve Fe- and Ir-related defect states. The resulting carriers are trap-derived conduction electrons which, although briefly entering the conduction band, are rapidly recaptured by the originating traps, leading to fast photocurrent decay without persistence [Figs. 5(c) and 5(d)]. A schematic summary of these excitation pathways and their roles in photocurrent dynamics is provided in Fig. 9.

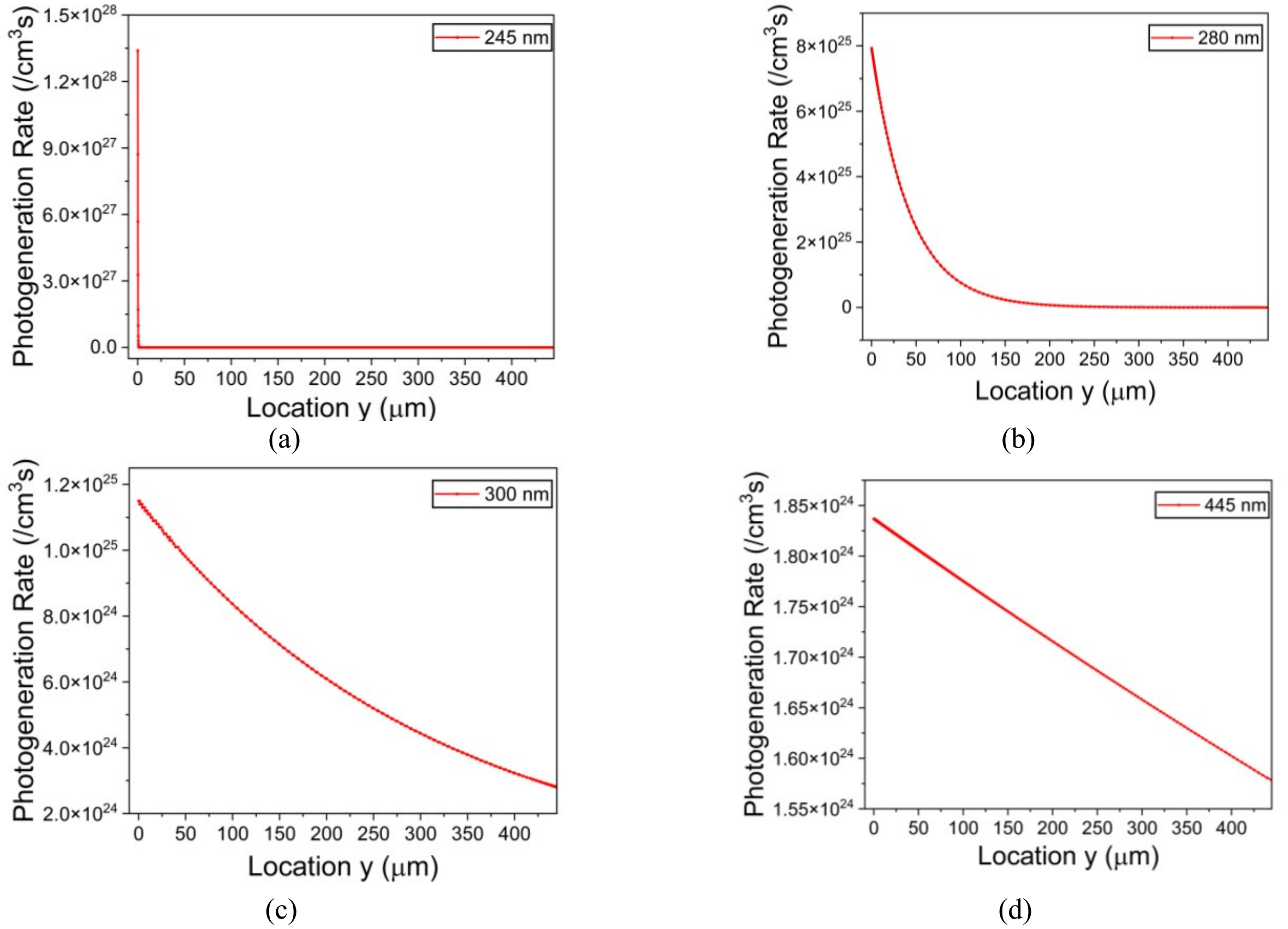

Fig. 8. TCAD Silvaco based simulated photogeneration rate as a function of device depth under different excitation wavelengths: (a) 245 nm, (b) 280 nm, (c) 300 nm, and (d) 445 nm.

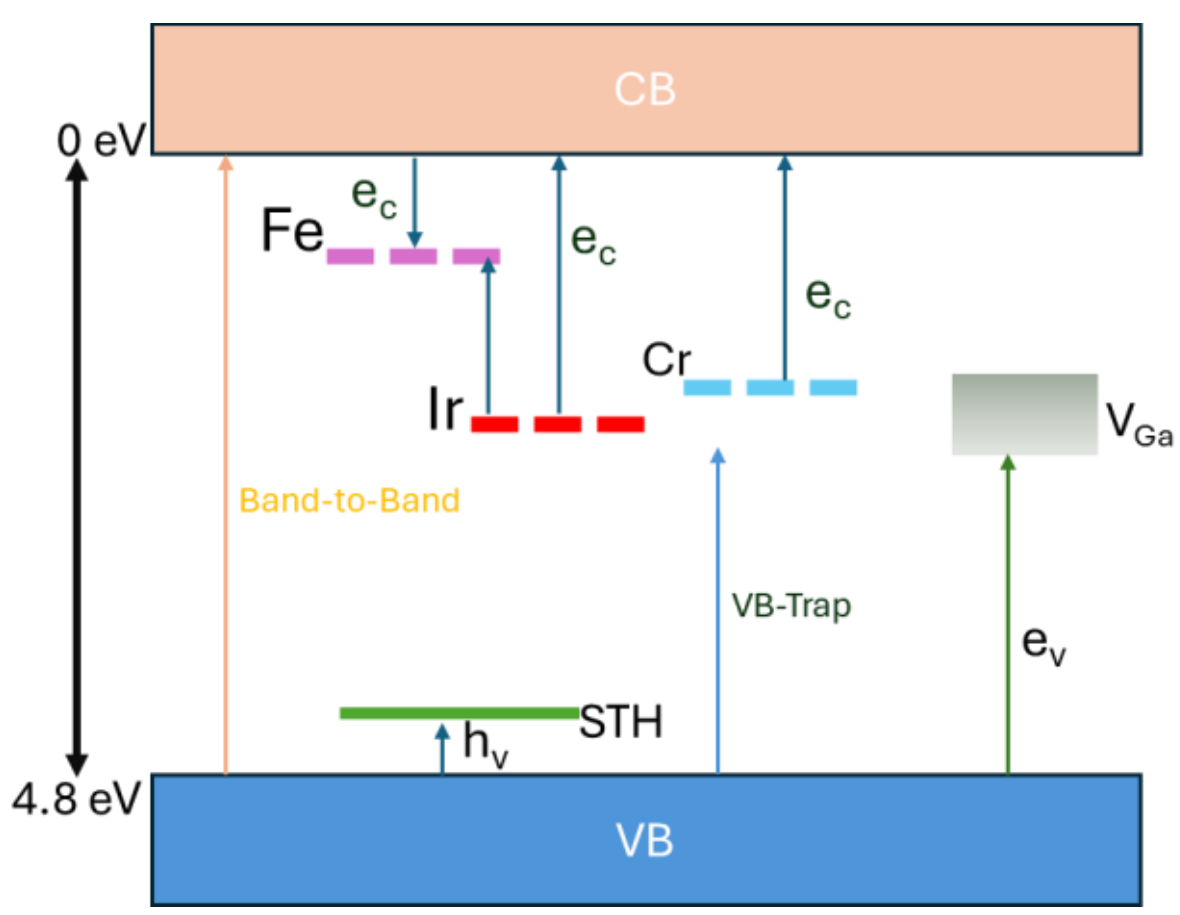

Fig. 9 Schematic band diagram of β-$Ga_2O_3$ highlighting the roles of deep levels and defect states in photocurrent dynamics.

TABLE II.
Comparison of normalized responsivity for lateral and vertical electrode PCSS structures fabricated with SiC, GaN, and $Ga_2O_3$ under intrinsic and extrinsic excitation wavelengths.

| **Structure** | **Material** | **Wavelength** | **Normalized responsivity (A-cm/W-kV)** | **Reference** |
|---|---|---|---|---|
| Lateral | SiC | 355nm, Intrinsic | $5.9 \times 10^{-6}$ | 19 |
| Lateral | SiC | 1064 nm, Extrinsic | $2.5 \times 10^{-7}$ | 19 |
| Lateral | SiC | 355 nm, Intrinsic | $1.97 \times 10^{-5}$ | 20 |
| Lateral | SiC | 532 nm, extrinsic | $9.74\times 10^{-7}$ | 21 |
| Vertical | SiC | 532 nm, extrinsic | $2.60 \times 10^{-6}$ | 21 |
| Vertical | SiC | 355 nm, Intrinsic | $1.78 \times 10^{-5}$ | 22 |
| Vertical | SiC | 532 nm, extrinsic | $1.25 \times 10^{-7}$ | 23 |
| Vertical | SiC | 532 nm, extrinsic | $2.17 \times 10^{-6}$ | 24 |
| Lateral | GaN | 355 nm, Intrinsic | $3 \times 10^{-6}$ | 25 |
| Lateral | GaN | 355 nm, Intrinsic | $1.52 \times 10^{-5}$ | 26 |
| Vertical | GaN | 532 nm, extrinsic | $1.96 \times 10^{-6}$ | 23 |
| Vertical | GaN | 532 nm, extrinsic | $3.6 \times 10^{-7}$ | 27 |
| Lateral | $Ga_2O_3$ | 245nm, Intrinsic | $8.2 \times 10^{-5}$ | This Work |
| Vertical | $Ga_2O_3$ | 300 nm, extrinsic | $4.01 \times 10^{-5}$ | |

A comparative analysis of the normalized responsivity for various PCSS structures is summarized in Table II. For SiC-based devices, intrinsic (above-bandgap) excitation at 355 nm consistently yields significantly higher responsivity than extrinsic (below-bandgap) excitation, irrespective of the electrode geometry. To the best of the author's knowledge, corresponding data for vertical GaN PCSSs operating under above-bandgap excitation has not been reported in the literature. In stark contrast to these materials, the vertical $Ga_2O_3$-based PCSS investigated in this work exhibits superior responsivity under below-bandgap excitation. This distinct behavior highlights the potential of $Ga_2O_3$ for high-responsivity PCSS applications that leverage extrinsic photoconduction.

## IV. CONCLUSION

In conclusion, this work has presented a systematic comparison of the photoresponse characteristics of lateral and vertical Fe-doped β-$Ga_2O_3$ PCSS devices under both intrinsic and extrinsic optical excitations. The devices were evaluated using above-bandgap (245 nm) and sub-bandgap (280 nm, 300 nm, and 445 nm) laser pulses to probe their respective carrier generation and collection dynamics.

Under 245 nm intrinsic excitation, the lateral PCSS demonstrated better performance due to the optimal overlap between strong surface absorption and the high surface electric field, resulting in a higher photocurrent and a faster response.

Conversely, under extrinsic excitation across the 280–445 nm range, the vertical PCSS consistently outperformed the lateral device. This is attributed to its uniform bulk electric field that enables efficient collection of deeply generated carriers.

Our results reveal a geometry-dependent performance crossover: lateral PCSSs achieve superior response under intrinsic excitation, whereas vertical PCSSs perform better under extrinsic excitation. This crossover, governed by absorption depth, carrier-generation profiles, and electric-field distribution, represents new insights specific to $Ga_2O_3$ and cannot be directly extrapolated from earlier SiC/GaN studies. For instance, in vertical SiC PCSS, Colt James *et al.* [22] reported peak photocurrent efficiency at 366 nm which is above-bandgap excitation. In contrast, the vertical $Ga_2O_3$-based PCSS examined in this work shows enhanced responsivity under below-bandgap excitation.

Overall, the findings in this paper offer critical insights for the design and optimization of $Ga_2O_3$-based PCSS architectures, guiding the development of cost-effective, high-performance switches tailored for specific optical triggering regimes.